\documentclass{article}
\usepackage{arxiv}
\usepackage[utf8]{inputenc} 
\usepackage[T1]{fontenc}    
\usepackage{hyperref}
\usepackage{ upgreek }
\usepackage{url}            
\usepackage{booktabs}       
\usepackage{amsfonts}       
\usepackage{nicefrac}       
\usepackage{microtype}      
\usepackage{lipsum}
\usepackage{float}
\usepackage{graphicx}
\usepackage{multirow}
\usepackage{array}
\usepackage{comment}
\usepackage{longtable}
\usepackage{ragged2e}
\usepackage{subfig}
\usepackage{amsmath}
\usepackage{longtable}
\usepackage{ amssymb }
\title{Few Shot Learning for Medical Imaging: A Comparative Analysis of Methodologies and Formal Mathematical Framework}

\author{
  Jannatul Nayem   \\
  Department of Electrical and Electronic Engineering\\
International Islamic University Chittagong\\
  Kumira, Sitakundu, Chittagong-4318, Bangladesh\\
  \texttt{jannatulnayem213@gmail.com} \\
   \And
    Sayed Sahriar Hasan   \\
  Department of Electrical and Electronic Engineering\\
International Islamic University Chittagong\\
  Kumira, Sitakundu, Chittagong-4318, Bangladesh\\
  \texttt{sayedsahriarhasan@gmail.com} \\
   \And
      Noshin Amina \\
   Department of Electrical and Electronic Engineering\\
  International Islamic University Chittagong\\
  Kumira, Sitakundu, Chittagong-4318, Bangladesh\\
  \texttt{noshinamina24@gmail.com} \\
   \And
Bristy Das\\
Department of Electrical and Electronic Engineering\\
International Islamic University Chittagong\\
Kumira, Sitakundu, Chittagong-4318, Bangladesh\\
\texttt{bristy44556@gmail.com} \\
   \And 
Md Shahin Ali\\
Department of Biomedical Engineering\\
Islamic University\\
Kushtia, 7003, Bangladesh\\
\texttt{shahinbme.iu@gmail.com}\\
   \And
  Md Manjurul Ahsan \\
  School of Industrial and Systems Engineering\\
  University of Oklahoma\\
  Norman, Oklahoma-73019 \\
  \texttt{ahsan@ou.edu} \\
   \And
 Shivakumar Raman \\
  School of Industrial and Systems Engineering\\
  University of Oklahoma\\
  Norman, Oklahoma-73019\\
  \texttt{raman@ou.edu}} 

\begin{document}
\maketitle

\begin{abstract}
Deep learning becomes an elevated context regarding disposing of many machine learning tasks and has shown a breakthrough upliftment to extract features from unstructured data. Though this flourishing context is developing in the medical image processing sector, scarcity of problem-dependent training data has become a larger issue in the way of easy application of deep learning in the medical sector. To unravel the confined data source, researchers have developed a model that can solve machine learning problems with fewer data called' Few shot learning’. Few hot learning algorithms determine to solve the data limitation problems by extracting the characteristics from a small dataset through classification and segmentation methods. In the medical sector, there is frequently a shortage of available datasets in respect of some confidential diseases. Therefore, Few shot learning gets the limelight in this data scarcity sector. In this chapter, the background and basic overview of a few shots of learning is represented. Henceforth, the classification of few-shot learning is described also. Even the paper shows a comparison of methodological approaches that are applied in medical image analysis over time. The current advancement in the implementation of few-shot learning concerning medical imaging is illustrated. The future scope of this domain in the medical imaging sector is further described.
\end{abstract}


\section{Introduction}\label{sec1}

The development and remunerative application of deep learning do not confine to computer science but also have flourished in other disciplines. Among them, one of the uplifting demeanors of deep learning systems is biomedicine. The deep learning algorithm is applied in various sub-sections of biomedicine like predicting images, quick disease detection, and biomedical image analysis~\cite{razzak2018deep, ahsan2020deep, ahsan2020covid}. The progress of deep learning becomes promoted in such a way that it gives better prediction than human-level accuracy regarding classification tasks. Medical image analysis requires a complicated understanding and detailed study of the extracted features of an image~\cite{ahsan2022machine, ahsan2021detection}. Even trained professionals find it quite tangled to understand. The manual process of recognizing the class of brain tumors or segmenting and annotating human organs requires years of experience, knowledge, and practice before someone can become an expert in this field. Nevertheless, the modern progress of deep learning systems makes it easier than before as it can extract features faster and learn from them to create an internal illustration of these features and make a prediction. The processing of the model, objective, and also high level of accuracy is always comparable with human accuracy~\cite{he2015delving, ahsan2023deep}.

It is evident that the application of deep learning is rapidly increasing in the medical sector. However, certain drawbacks become an obstacle in the way of proper illustration of deep learning in the sector of medical image processing. Among them, scarcity of labeled data and class inconsistency is considered one of the biggest dependencies of the deep learning process. The deep learning model's accuracy and precision largely depend upon a large number of the training dataset~\cite{ahsan2022machine}. This issue might not create a big problem for many medical cases, but there are some rare cases where a large number of data might not be available for its confidentiality. It is hard to find available data for some rare cases which ultimately do not provide satisfactory results regarding deep learning image analysis for the medical sector. Even though there is no surety that the acquiring data would be enforceable and enduring to the model while training the dataset to solve the data dependency problem of deep learning has to be solved in a human mind processing way with less data~\cite{lee2017deep}.

Few-shot learning is the developed version of deep learning with limited data sources to generalize entire classes of data. This method works by extracting raw data features for differentiating the classes. Convolutional neural networks fully depend on the depiction of a larger training dataset. The problem of the general representation of convolutional network neural network turns out to be a specific illustration with the help of few-shot learning. In the following chapter, there will be a discussion and illustration of the overview of few-shot learning and a comparison of methodological analysis of few-shot learning approaches~\cite{litjens2017survey}. Few-shot learning is a model based on deep learning systems with fewer datasets applied for training a model to make a prediction and classification based on data analysis. This process of training the model is different from the conventional process for simplifying the model with higher prediction accuracy. Normally, a model trained belongs to the previous class. The model consists of a support set with n number of examples along with k various unseen classes. The performance of few-shot learning determines which support set sample is associated with the query set samples~\cite{shen2017deep}.

To complete the task, few-shot learning is used to develop certain models for medical image classification. AffinityNet is one of the models that exhibits outstanding performance on disease-type prediction by extracting local abstract features. Another approach, the Siamese Network, is demonstrated for retrieving pictures from medical datasets. The Triplet Network is a developed version of the Siamese Network that identifies the modality of brain imaging. Domain generalization is another tackling point in a few-shot learning algorithm, especially for high-dimensional non-linear biomedical images. Covariate shift is a tried statistical method for domain adaptation and generalization for this sector. The gradient-based meta-learning domain generalization agnostic method is used to improve the generalization model for the unseen dataset for three domains: new anatomies, diseased population, and unseen anatomies. It is validated by computed tomography (CT) vertebrae segmentation task~\cite{khandelwal2020domain}.

Few shot learning pathways are widening because of less reliability of training data. Another significant challenge in the medical sector is to track gastrointestinal issues, the kidney, and the liver through Endoscopy. Nevertheless, large datasets from different clinical sources can create an extensive label bias, making the model unusable. But a novel additive angular margin metric model, which is a modified framework of a prototypical network, is introduced for classifying endoscopy test samples from an unseen dataset that is trained with less training data~\cite{ali2020additive}. Even the gradient-based Reptile and distance metric-based Prototypical networks, few-shot learning techniques, are applied to identify skin lesion image datasets. The proposed method is named after the Meta-Derm Diagnosis network. Here, the problem is classified into 1-shot, 3-shot, and 5-shot classification problems and used G-convolutions to rectify the network's performance~\cite{mahajan2020meta}. The requirement of limited data makes few-shot learning limelight in this deep learning era. New dimensions of few-shot learning applications like ophthalmology are initiated through the practice of few-shot learning. Though deep learning is applied in pathological clinical cases, there are some rare cases like papilledema or anterior ischemic optic neuropathy are failed to conduct a few shot learning methods due to the scarcity of training datasets. Convolutional neural networks (CNNs) with the t-SNE visualization tool are incorporated within a ROC curve (AUC) greater than 0. whose output is better than comparable other frameworks~\cite{quellec2020automatic}.

In enhancing clinical diagnosis and treatment planning, image segmentation is an inevitable task to annotate a huge amount of training datasets. A few shot segmentation was proposed with a new furnished structure to solve the dilemma. Medical Prior-based Few shot Learning Network (MPrNet) and Interactive Learning-based Test Time Optimization Algorithm (IL-TTOA) for supporting query image and interactive purposes to alleviate the hassle of medical experts and to test the result of this model with the real experimentation of human clinical dataset~\cite{feng2021interactive}. Therefore, another segmentation model called global correlation network with discriminative embedding (GCN-DE) is proposed to fasten generalize unseen datasets with spatial consistency and regularity. This method is applicable to alleviate computational complexity by creating convolutional layers between support and query images in feature space~\cite{sun2022few}. Furthermore, image segmentation through few-shot learning is applied comprehensively in skin and polyp clinical datasets with implicit model agnostic meta-learning iMAML algorithm, which embodied attention-UNet mechanism and conjugate gradient for attaining expected results~\cite{khadka2022meta}. The revolutionary contribution in image segmentation continued through the research of 3D medical image segmentation where U-Net network, a model to calculate the relationship between 2D support and query dataset through bidirectional GRU modules and transfer learning technique for the betterment of performance. The model was tested on various internal and external organ datasets~\cite{kim2021bidirectional}.

In the following chapter, there will be a discussion and illustration of the overview of few-shot learning and a comparison of methodological analysis of few-shot learning approaches. The chapter is designed into eight main parts. The first two-part are about the introduction and existing work with sufficient references. The third part is about the overview of a few shot learning where the terminological aspects are described. The fourth part is about the classification subsection of the few-shot learning method. The fifth part explains the formal mathematical expression, and the sixth part compares the methodological approaches of few-shot learning in medical imaging. In the last two sections, the future scope of medical imaging and the chapter overview is described in the discussion part.

\section{Related Work}

Due to the need for only a small number of training data, few-shot learning attracts the most interest. There is immense research on the advancement of few-shot learning in the medical sector. Recently Farkaš et al (2020) proposed a Context Adaptive Metric Model (CAMM) with an adaptive ability consisting of two parts Context parameter module, which accords the characteristics of the feature extractor, and a Self-evaluation module that fine-tuned context parameters~\cite{wang2020context}. This model can be used for extracting key features for most of the metric models to solve the overfitting problem in few-shot learning. Besides, another research on ensemble-based deep metric learning (EBDM) consists of the shared part, which is used to cut short the parameter numbers and the exclusive part for the feature extraction network in each learner. This model can be used for image labeling for a new class with higher accuracy~\cite{zhou2020ensemble}. Therefore, Li et al (2020) introduced a More Attentional Deep Nearest Neighbor Neural Network (MADN4) with local descriptors and an attention mechanism combining Convolutional Block Attention Module (CBAM) to deal with noise problems along with local feature extraction~\cite{li2020more}. Few-shot learning (FSL) is used practically to distinguish between Chronic Obstructive Pulmonary Disease (COPD) patients and asthma subjects with limited data, where Siamese-based zero-shot, one-shot, and five-shot learning models are implemented for the dissimilation of COPD patients from asthma~\cite{zarrin2020implementation}.

For adaptation from relatively few samples in the global feature technique, few-shot learning is an acknowledged concept in the medical field. Early disease detection in the medical sector is a crucial challenge nowadays. That’s why the application of Few shot learning is increasing daily in medical image classification~\cite{munkhdalai2017meta}. In addition, Ma et al (2018) Affinity Net with the k-Nearest-Neighbor attention pooling layer based on semi-supervised few-shot learning exhibit outstanding output on disease type prediction by extracting abstract local features~\cite{ma2019affinitynet}. The research was continued through Siamese Network, a first-rate network~\cite{chung2017learning} which is used to recover medical images from datasets, and Triplet Network, a more advanced Siamese Network used to recognize the brain imaging modality~\cite{puch2019few}. Even few-shot learning is applied in Glaucoma diagnosis successfully~\cite{kim2017few}. Also, the classification of diabetic retinopathy has been improvised through meta-learning~\cite{hu2018meta}.

Over the last few years, few-shot learning has focused on medical image segmentation, where many techniques have been initiated. One of the works from this domain where Ronneberger et al~\cite{ronneberger2017invited} performed medical image segmentation by U-Net with great performance. The segmentation of the condition of tissue by color deconvolution deep neural networks have been performed by Lahiani et al~\cite{lahiani2019generalising}. Press and excite blocks have been used to segment volumetric images~\cite{roy2020squeeze}. Therefore, Generative Adversarial Networks~\cite{goodfellow2020generative}, 3D volumes of multiple modalities~\cite{mondal2018few}, and data-augmented networks~\cite{zhao2019data} have been applied for preparing and executing one-shot biomedical image segmentation. Zhongqi et al~\cite{yue2020interventional} developed an Interventional Shot Learning (IFSL) with three feasible IFSL algorithmic effectuation based on backdoor adjustments for orthogonal contribution with existing fine-tuning and meta-learning FSL methods.

Researchers found new dimensions for implementing few-shot learning in real-life projects with a broader sense. They represented their new methodology of few-shot learning to solve the segmentation problem based on medical cases. Rabindra et al.~\cite{khadka2022meta} proposed an optimization-based implicit model agnostic meta-learning (iMAML) algorithm for segmenting medical images by analyzing unseen datasets with high accuracy. It generates a generalization method for training skin and polyp datasets which outperform over supervised baseline model and two few-shot segmentation approaches by over 2\%–4\% in dice score. In~\cite{guo2020broader}, the Broader Study of the Cross-Domain Few-Shot Learning (BSCD-FSL) model has been proposed with a varied repository of image data, including crop disease images, satellite images, dermatology images, and radiology images. The state-of-art meta-learning method performs better than the conventional meta-learning process. A Semi-supervised few-shot learning model for medical image segmentation where few unlabeled images are trained in an episodic process to get efficient feature extraction. The model is generalized in skin lesion segmentation datasets~\cite{li2021supervised}.

Additionally, the study of few-shot learning using a prototypical network reached a new level. A model known as the additive angular metric has been introduced as a framework for a typical network. It helps to identify between multi-modal, multi-center, and multi-organ gastrointestinal endoscopic unobserved data. It is used to assume the label class of test samples, and the suggested model outperformed other models by roughly 7\%~\cite{khandelwal2020domain}. Conv-4 backbone with different backbones represented to implement the model. In addition, the research work~\cite{sun2022few} illustrated a domain generalization model with an agnostic meta-learning framework to improve unseen test distribution by applying a computed tomography (CT) vertebrae segmentation approach. Shot learning is employed in this generalized method with fewer unknown datasets. The term "global correlation network" was used to describe an image segmentation technique that quickly captures previously unknown classes between a support and query image~\cite{vo2023neural}. Moreover, it experimented with CT and MRI modalities and discriminative embedding loss. 

In~\cite{kim2021bidirectional}, Neural Architecture Search (NAS), a network architecture, has proposed to solve the manual trial and-error processes issues during extracting features of unstructured data. The network developed computer vision applications and ennobled the validity of image classification and segmentation.
The scarcity of medical data makes few-shot learning a prominent research topic in the deep learning domain, and the research is flourishing daily. Using the 3D image segmentation model~\cite{kim2021bidirectional}, U-Net created a connection between the data and a query image of 2D slices. In addition to this, the features of adjacent slices are extracted using a bidirectional gated recurrent united (GRU). In addition, transfer learning is used to identify the targeted image's characteristics. Three procedures are employed~\cite{feng2019few} to establish few-shot human activity recognition (FSHAR): deep learning model training on source domain samples, cross-domain relevance on source domain samples and target training samples, and finally, fine-tuning to initialize the targeted model. The Implicit Model Agnostic Meta-Learning (iMAML) model is employed in the segmentation of medical images to address the problem of generalizing previously unknown data~\cite{khadka2022meta}. It may apply practically to skin and polyp datasets by learning from hidden training samples. Medical picture segmentation problems with annotation are resolved with Interactive Few-shot Learning (IFSL)~\cite{feng2021interactive}. It was created by Medical Prior-based Shots Learning Network and Interactive Learning-based Test Time Optimization Algorithm (IL-TTOA). Without prior training, it can support the requested image and optimize the segmentation models.

\section{Overview ff Few Shots Learning}

Few shot learning, an illustration of meta-learning, and a structural representation of the machine learning model are applied to categorize the new data based on a few labeled datasets with its pre-trained model. Earlier, computer vision aimed to be the only focused research domain of FSL. Still, it is progressing with the NLP but slower due to the scarcity of natural language data and assembled benchmarks.
Since domain-related technicalities, expressions, and the coalition is arousing difficulties, the Application of few-shot learning in medical images is in the limelight of attention. Here, a small number of data are trained with prior knowledge before generalizing the model in an efficacy manner. Afterward, the data are further proceeded for fine-tuning. The prior knowledge is applied in three processes; to augment training data, to suppress hypothesis space, and to attain parameters.

\subsection{Important Terms of Few-Shot Learning}

\begin{figure}[ht]
    \centering
    \includegraphics[width=.8\textwidth]{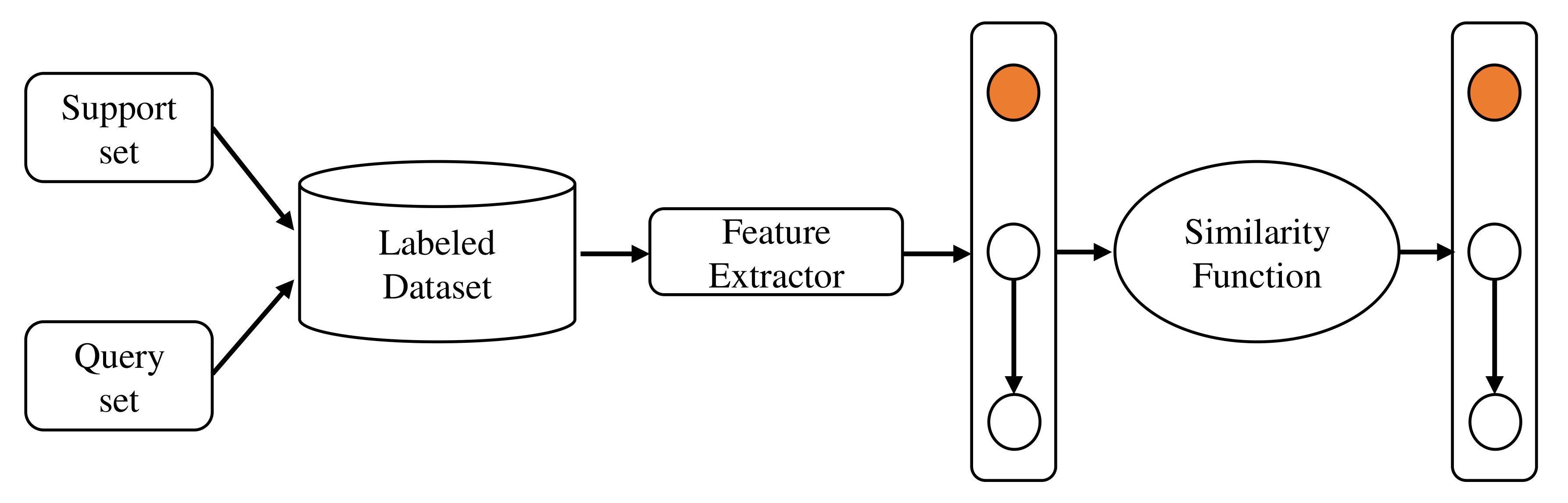}
    \caption{A simple illustration of a few-shot learning model containing a support set and query set where the features are extracted from both datasets and made decision-based on a similarity function.}
    \label{fig1}
\end{figure}

To understand the concept of few-shot learning, consider two sets (C train $\cap$ C test= $\phi$) where C train contains a larger amount of training dataset whereas ($X_{train}^{N}$, $Y_{train}^{N}$) whereas C test has comparatively less training dataset. From this dataset, the analogies of the medical domain can be calculated by studying class categories related to clinical patterns. No algorithm can perform more effectively than few-shot learning when it comes to generalizing a model with fewer datasets. In FSL, n is the number of classes the pre-trained model can generalize, and k is the number of labeled samples that are part of the n number of support sets. The more the number of 'Values, the more difficult the task is, whereas the less the number of k values, the more difficult the task is. Moreover, one-shot learning is shown when k=1. Zero-shot learning is the case when k=0.The training set for the model consists of k classes and n training examples, where n << N is a small sample of training examples drawn from the training set of classes n $\times$ k $\ll$ $|C train|$. They are defined as the Sb support set subset in Fig. \ref{fig1}. On the other hand, there are other instances outside support set Sb, such as query set Q b, which uses the support set to gather information and feature extraction for generalization. It is composed of recent and conventional data. These two sources of data are equally important for validating the examples of supervised learning algorithms. As a result, altering the query set Qb depends on modifying the support set S b. The following model parameters $\phi$ should be learned in order to estimate the query set f$\phi$: RD $\rightarrow$ Rd (D $\gg$ d)~\cite{ali2020additive}.

\section{Few Shot Learning Classification-Based Algorithm}

\begin{figure}[ht]
    \centering
    \includegraphics[width=.8\textwidth]{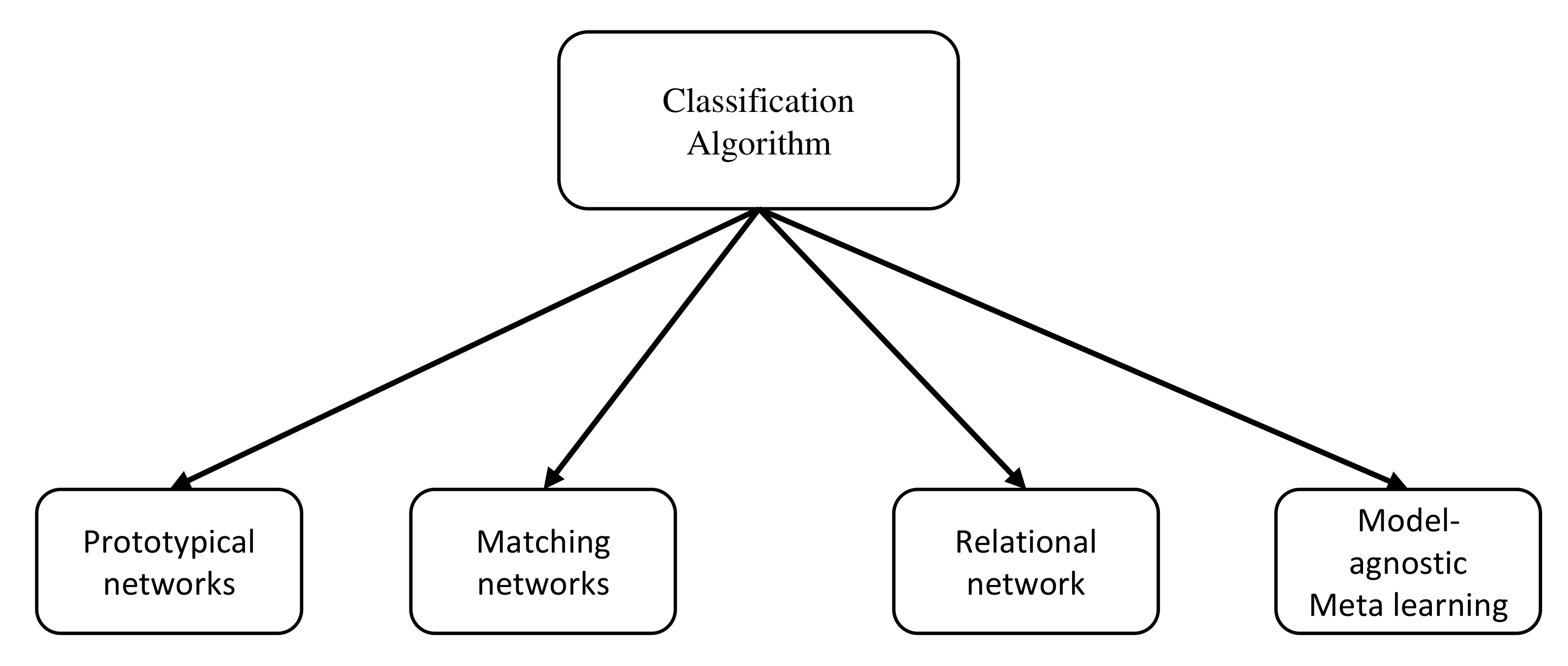}
    \caption{A diagram of a few shot classification-based algorithms representing Prototypical networks, Matching networks, Relational networks, and Model-agnostic Meta-learning.}
    \label{fig2}
\end{figure}

Fig. \ref{fig2} shows that categorizing medical images using few-shot learning is a growth-oriented area. The classification process divides inputs into groups based on their differences. The labeled data is categorized using a similarity method, producing accurate results. With a smaller sample, this technique concentrates on locating and retrieving feature data. The K-shot N-way approach of few-shot categorization makes use of fewer data.

\subsection{Prototypical Networks}

Prototypical networks~\cite{yue2020interventional} illustrate the concept of a single prototype made of multiple points which represent each class. For instance, a neural network is trained with a car image and without a car image for binary classification. As the network includes images without a car, many unnecessary features are to be processed within a large range that consumes computing consumes, computing energy and time. To solve the problem, the network considers only the positive classes rather than the negative classes. Prototypical networks operate in this incantation where the prototype is constructed with the most momentous characteristics of the positive classes that are matchable with unknown and unseen samples in the future. An internal low-dimensional embedding of discrete data is made in the whole class. An embedding in a neural network means containing a large number of car images that are elucidated as a single vector. This method solves the categorical variables issue and resembles similar entities in a vector space. This embedding method is also used to lessen the loss of the task. The equation illustrates the prototype:

\begin{equation}
L_k = \frac{1}{S_b^k} \sum_{(X_i, Y_i) \in S_b^k} f_\phi(X_i)
\end{equation}

\begin{equation}
d (f_\phi(X) , L_k =
\arccos \frac{f_\phi(x) \cdot L_k}{f_{\oplus}(x) \cdot L_k}
\end{equation}

\begin{equation}
p_{\emptyset}\left(y_k \mid X\right)=\frac{e^{\cos \left(d\left(f_\phi(X) L_k\right)\right)}}{\sum_{k^{\prime}} e^{\cos \left(d\left(f_\phi(X), L_k\right)\right)}}, \quad k^{\prime} \in[1, \ldots, k]
\end{equation}

\begin{equation}
J_\varnothing = -\log p_\varnothing(\langle y_k \rangle  X)
\end{equation}

The network calculates the $M$-dimensional vector $L_k$, where $\phi \in \mathbb{R}^d$ is a prototype representation, $x_i$ is a $d$-dimensional feature vector, and $Y_i$ is the corresponding label for support set $S_b$ as well as for every class $k$ (see Eq. (1)). Afterward, it determines the distance $d(\cdot)$ between the query example $X$ and the prototype representations $\phi$ of any class $k$ (see Eq. (2)). It provides probability in accordance with Eq. (3). After evaluating two mostly used distance metrics: Euclidean and cosine, cosine metric turns out to be more effective in continuous disease progression. Therefore, for the query example $X \in Q_b$, the parameters are computed using stochastic gradient descent (SGD) to reduce the average negative log probability (see Eq. (4)) of the target label $y_k$.

\begin{figure}[ht]
    \centering
    \includegraphics[width=.65\textwidth]{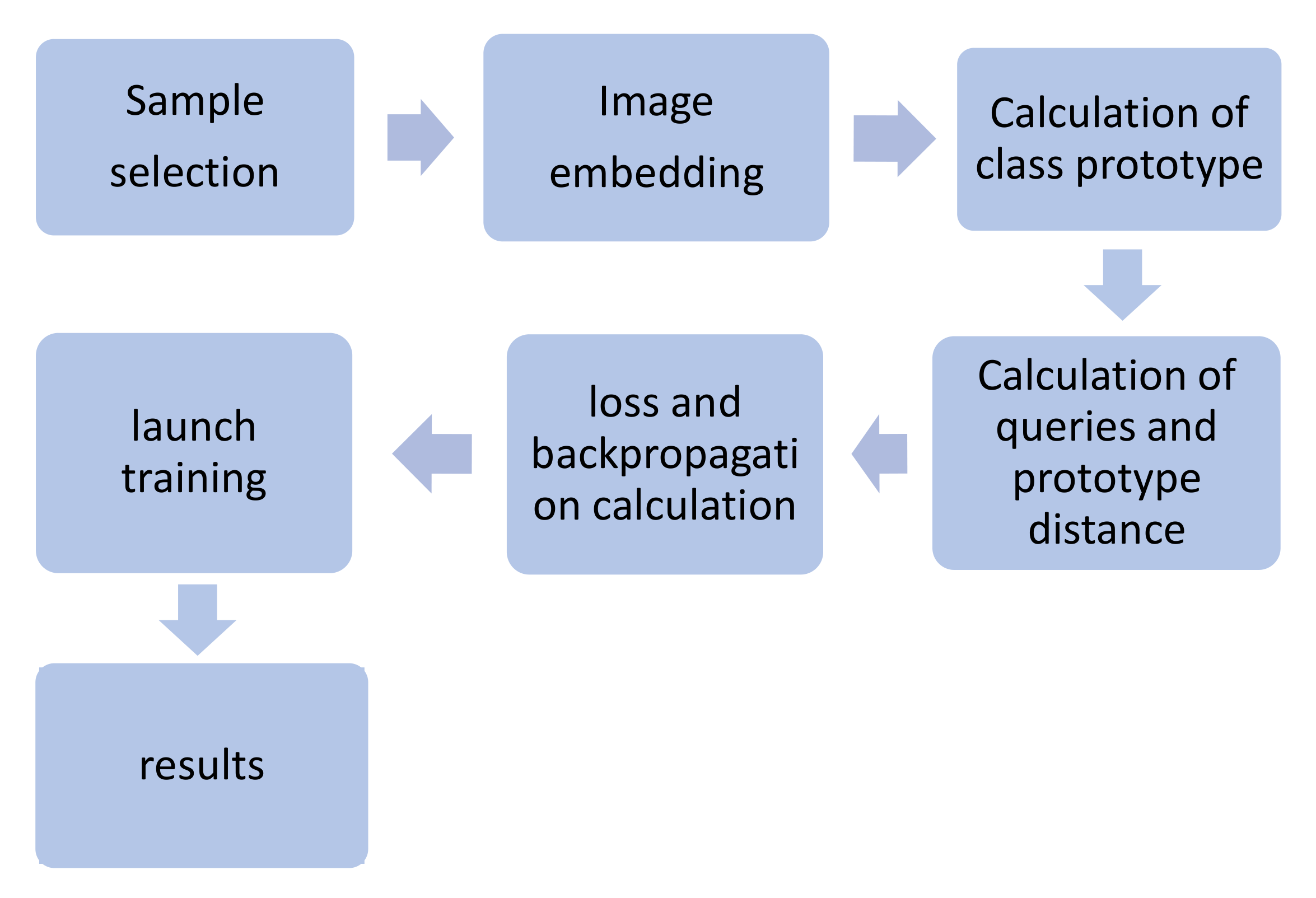}
    \caption{A representation of the flowchart of a prototypical network, after selecting the sample, the class prototype, the distance between query and prototype, loss and propagation are calculated, and the launch training. }
    \label{fig3}
\end{figure}

Fig. \ref{fig3} shows that after training the network, the query of unseen samples is compared with the internal embedding of the class. Then the attributes are dimensionally reduced and compared with the similarity function for further results.

\subsection{Matching Networks}

Matching networks~\cite{santoro2016meta} is another classification model which endeavors to learn the working of a classifier using fewer data with the help of memory augmentation. This network assimilates the $k$-nearest neighbor's algorithm, which needs no training dataset, and a conventional neural network that needs thousands of samples to learn the model. The equation illustrates an attention mechanism $\alpha$, which compares $\hat{x}$ and $x_i$ for accessing memory.

To establish the matching network, the input image $x_i$ and their corresponding label $y_i$ are represented as a support set $S={(x_i,y_i)}_{i=1}^k$. Then a classifier, $c_s(x)$, maps the support set. The probability of a new example $\hat{x}$ in this network falling into a certain class of labels must be estimated, and this likelihood is represented by $P(\hat{y} \mid \hat{x},S)$.

The problem can be mathematically formulated as determining the output category with the maximum probability (see Eq. (5)). The calculated output is defined by the given equation:

\begin{equation}
\hat{y} = \sum_{i=1}^{k} a(\hat{x},x_i)y_i
\end{equation}

\begin{figure}[ht]
    \centering
    \includegraphics[width=.8\textwidth]{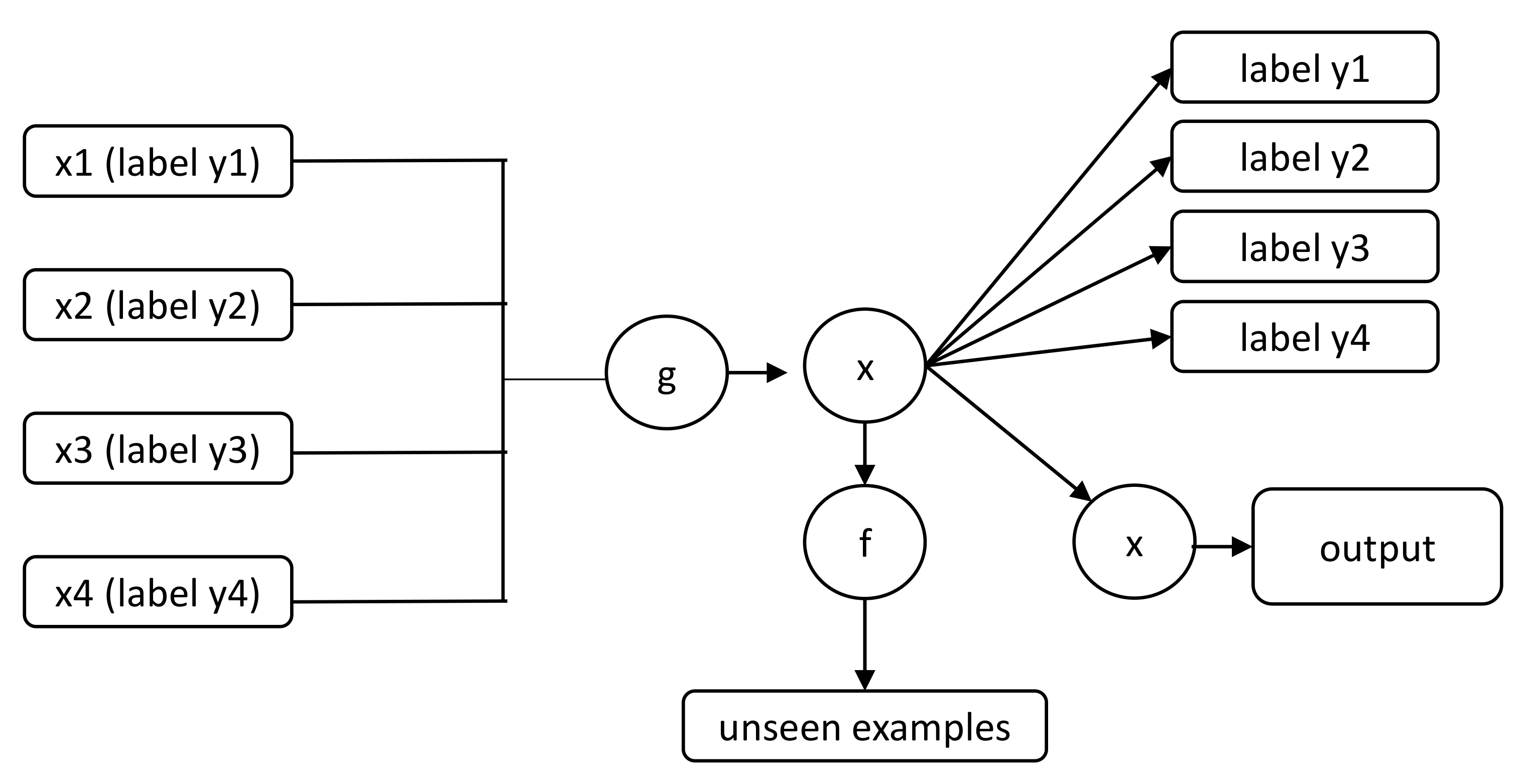}
    \caption{The Block diagram of the matching network where the support set is fed into an embedding function g for mapping and attention mechanism x for identifying the unseen examples and finding the probability function f of the matching rates with the known label.}
    \label{fig4}
\end{figure}

Therefore, the attention mechanism basically measures the cosine distance between support set examples and new examples. Hence, the process goes through the SoftMax function for normalizing the output in the range between 0 to 1. Then the support set passes through an embedding function g, generally a bidirectional LSTM for mapping the support set. The encoding function g embeds the full set S based on other elements of the set and each individual xi and becomes g (xi, S). Suppose a small batch B is considered when the training examples are few. Generally, the matching network extracts the data from the given support set S by encoding function f and reduces error, as shown in Fig. \ref{fig4}.

\subsection{Relational Networks}
Relational networks (RNs)~\cite{zhang2020deep}, a sub-section of a neural network, deals with relational reasoning with the help of plug-in components. The construction of this network is to co-relate between two input samples based on reasoning. The function that is used for the relational network is given below:

\begin{equation}
RN(O) = f_{\phi} \sum_{i,j} g_{\theta} (O_{i,} O_j)
\end{equation}

\begin{figure}[ht]
    \centering
    \includegraphics[width=.75\textwidth]{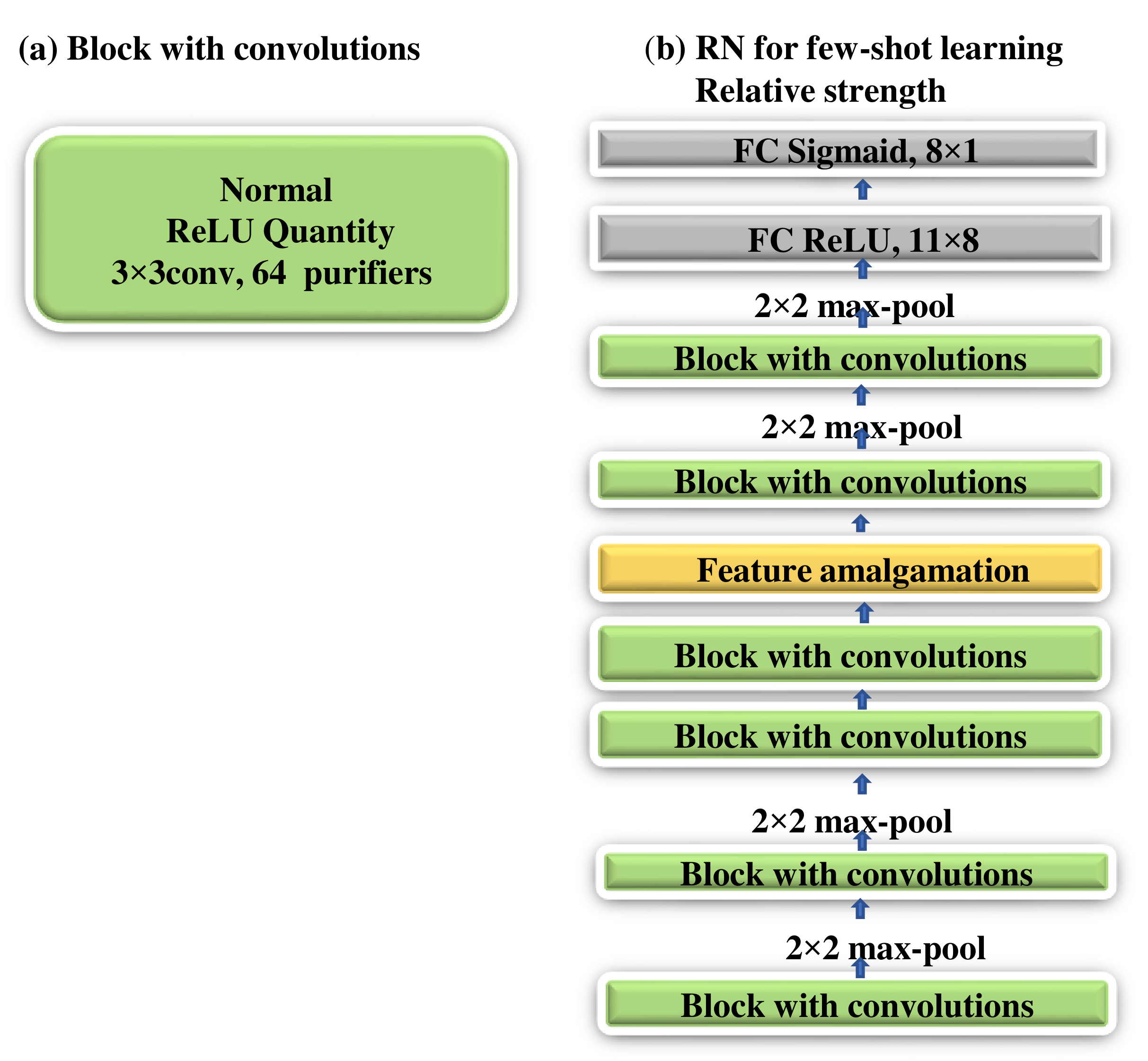}
    \caption{an illustration of a relation network framework that incorporates six convolutional blocks for embedding purposes (a) in few-shot learning (b) }
    \label{fig5}
\end{figure}

O is considered a set of objects used for function f$_{\phi}$ and g$_{\theta}$. The function g$_{\theta}$ determines the relationship between object pairs. Suppose a traditional Multi-Layer Perceptron model is applied to extract all the objects from the input object pair only once by conducting dimensionality reduction through the complexity n$^{2}$. Another significant characteristic is that it neither depends on input nor output order. This formulates the relation between the objects. Only images or only text do not define a relational network. This network only pursues the input that is in the embedding form. The illustrating object semantics can be neglected to make the model dynamic.
Relation network work regarding the distance metrics for performing classification. Generally, fixed metric (Euclidean or cosine) and conventional metric((linear) Mahalanobis metric) are used for fixed feature representation. Still, the Relation network is trained to learn deep embedding and non-linear metrics through similarity functions. This is particularly useful because fixed metric works on the basis of linear separability after the embedding and emphases choosing fixed metrics like Euclidean, cosine, or Mahalanobis. In contrast, the Relation network develops a non-linear similarity metric which is better at embedding and classifying matching/mismatching pairs. And this network executes a flexible function approximator for the efficacy of the model.
As the majority of few-shot learning models employ four convolutional blocks for the embedding module, the network adopts the same architectural configuration for a fair assessment; see Fig. \ref{fig5}. To be specific, each convolutional block contains some particular layers named batch normalization layer, a ReLU nonlinearity layer, and a 64-filter 3$\times$3 convolution. A 2$\times$2 max-pooling layer is likewise present in the first two blocks but not in the last two. As a result, the network can be employed the output feature maps for the relation module's additional convolutional layers. Furthermore, two convolutional blocks and two fully connected layers comprise the relation module where 3$\times$3 convolution with 64 filters for an individual convolutional block along with batch normalization, ReLU non-linearity, and 2$\times$2 max-pooling. In order to maintain an acceptable range for the relation score, all connected layers are ReLU apart from the output layer.

\subsection{Model-Agnostic Meta-Learning}
Model-Agnostic Meta-Learning~\cite{finn2017model} model is eminent for its supple translation and solution for a comprehensive range of problems. When data availability increases, the method gains whippy adaptability during execution. The meta-learning method trains a small data set in such a way that it can eliminate over-fitting. The actual objective of this method is to reduce the loss of functions for newly assigned tasks. Suppose $\theta$ is the parameter vector used for an individual task selected at the time of the training phase from the support set. Afterward, the network is trained with a few swatches to generate feedback for that particular task. Subsequently, a test set is exercised to evaluate the model’s performance as well as its parameters. Furthermore, f$_{\theta}$  defines the parameterized function associated with the parameter vector $\theta$. The parameter vector $\theta$ changes to $\theta'_{i}$ when the model is loaded with a new task T$_{i}$. Hence, This task is done by fine-tuned parameter where $\alpha$ defines the rate of learning, L illustrates the  loss function of T$_{i}$ task, and f$_{\theta}$ represents the model (see Eq. 7). Henceforth, the batch-wise format is followed in this model with a few numbers of examples. Equation 1 for this model is given by,

\begin{equation}
\theta'i = \theta - \alpha \delta{\theta} L_{T_i}(f_{\theta})
\end{equation}

In addition, Model-Agnostic Meta-Learning is applied in the domain of few-shot learning to implement a new function of supervised task from the given previous similar tasks. The model is applicable regarding few-shot regression and classification. Regression simply requires a few data points from a continuous-valued function to determine the result. Prior to that, comparable statistical features are input into the model. In the case of supervised regression and classification problems, the model considers the horizon $H=1$, timestep subscript on $x_{(t)}$ as it works with the single input and output. The task $T_i$ propagates $K$ i.i.d. observations $x$ from $q_i$, and the error between the output of $x$ and its corresponding target value $y$ is considered the task loss. To find the output of the loss function, cross-entropy is used for supervised classification (see Eq. 8), and mean-squared error (MSE) is used for regression as shown in Eq. (9) where $x^{(j)}$ defines the input pair sample and $y^{(j)}$ illustrates the output pair sample. The loss function of the regression model is given below:

\begin{equation}
L_{T_i}(f_{\theta}) = \frac{1}{K}\sum_{j=1}^{K}(f_{\theta}(x^{(j)}) - y^{(j)})^2
\end{equation}

Correspondingly, the cross-entropy loss for discrete classification tasks has the following form: 
\begin{equation}
L_{T_i}(f_\varphi)=\sum_{(x^{(j)},y^{(j)})\sim T_i} y^{(j)} \log f_\varphi(x^{(j)}) + (1-y^{(j)}) \log (1-f_\varphi(x^{(j)}))
\end{equation}

\section{Formal Mathematical Statements of Few Shot Learning Problems}

The most intriguing challenge relates to how AI models generalize the useful training data and memorize label data, in addition to how they learn from training examples. The deep learning model is more efficient as a result of the continual process of obtaining the answers to these centralized inquiries. Few-shot learning has recently emerged as one of the trendy topics in the field of deep learning~\cite{sejnowski2020unreasonable}. With a few examples of training data, several algorithms, and models are demonstrated to be useful for greater generalization. This paper will describe the mathematical underpinnings of a few generalized few-shot learning problems and their solutions. To build up a concrete mathematical foundation, the generalized problems of few-shot learning are categorized into two parts: learning from new examples or learning a new class from less dataset.

\subsection{Problem 1(learning from new examples)}

Let $U$, $U\subset\mathbb{R}^d$ where $U$ defines the input object modeling (see Eq. 1). 

\begin{equation}
F: U \rightarrow L
\end{equation}

where $F$ is a classifier and further illustrated $f: U\rightarrow X$, $X\subset \mathbb{R}^n$, $X$ is the latent space of classifier. And let $U_{new}={u_{1},\ldots,u_{k}}$, $k\in \mathbb{N}$; $\mathbb{N}$ is the natural number, $u_{i}\in U_{new}$ defines a new finite set of classifier $F$. Let $l_{new}\in L$ represents a label of a new object $U_{new}$. Therefore, $p_{e}$ is denoted for a positive number in the $(0,1]$ range.
Discovering a method $A(U_{new})$ that generates a function $g^{*}:X\rightarrow L$ such that for $u$ taken from the distribution $P_{u}$ (see Eq. 11 and Eq. 12):

\begin{equation}
\mathrm{P}(F[g^* \circ f(u)] = l_{\text{new}} \text{ for all } u \in U_{\text{new}}) \geq p_n
\end{equation}

And

\begin{equation}
P(F[g^* \circ f(u)] = F[g \circ f(u)]) \geq p_e
\end{equation}

\subsection{Solution of problem 1 with theorem}

From Eq 10, classifier $F$ is defined, and let $U_{\text{new}}={u_{1},\ldots,u_{k}}$, $k\in\mathbb{N}$, $u_{i}\in U_{\text{new}}$ defines a new finite set of classifier $F$. Let $l_{\text{new}}\in L$ represent a label of the new object connected to new elements.

Let $Y={x_{1},\ldots,x_{k}}$, where $x_i=f(u_i)$, $i=1,\ldots,k$, is defined as the explanation of the set $U_{\text{new}}$ in the classifier's latent space.

\begin{equation}
\bar{x} = \frac{1}{k} \sum_{i=1}^{k} x_i
\end{equation}

Being the representation's practical average,
\begin{equation}
(x_{\bar{}}, x_i) \geq 0 \quad \text{for all } x_i \in Y
\end{equation}

The outline is,
\begin{equation}
g^*: g^*(x)= \begin{cases}\ell_{\text {new }}, & \text { if }\left(\frac{\bar{x}}{\|\bar{x}\|}, x\right)-\theta \geq 0 \\ g(x), & \text { otherwise }\end{cases}
\end{equation}

Calculated by,

\begin{equation}
\theta=\min_{i\in{1,\ldots,k}}{(\frac{\bar{x}}{\bar{x}},x_i)}
\end{equation}

is a means of solving Problem 1,

\begin{equation}
p_e = 1 - C_x/2 [r(1 - \theta^2)^{1/2}]^n
\end{equation}

Where C$_x$ is a constant.

\subsection{Problem 2 (learning from new class)}

From Eq. 10, classifier $F$ is defined and let $U_{\text{new}}={u_1,\dots,u_k}$, where $k\in \mathbb{N}$, $u_i \in U_{\text{new}}$ is a finite independent sample from a distribution that is uniformly distributed $P_{\text{new}}$ and $l_{\text{new}}\in L$ represents a new label for every individual element from $P_{\text{new}}$. Therefore, $p_e$ and $p_n$ are denoted as positive numbers in the $(0,1]$ range. Discovering a method $A(U_{\text{new}})$ that generates a function $g^*:X\rightarrow L$ such that for $u$ taken from the distribution $P_{\text{new}}$, see Eq. 18 and Eq. 19.

\begin{equation}
P(F[g^* \circ f(u)]=l_{new}) \geq p_n
\end{equation}

\begin{equation}
P\left[F[g^*\circ f(u)]=F[g\circ f(u)]\right] \geq p_e
\end{equation}

\subsection{ Solution of problem 2 with theorem}

From Equation 10, the classifier $F$ is defined. Let $U_{new}={u_{1},\ldots,u_{k}}$, where $k\in\mathbb{N}$ and $u_{i}\in U_{new}$, be a finite independent sample from a distribution that is uniformly distributed $P_{new}$, and let $l_{new}\in L$ represent a new level to every individual element from $P_{new}$~\cite{tyukin2019high}. Let $Y={x_{1},\ldots,x_{k}}$, where $x_{i}=f(u_{i})$ for $i=1,\ldots,k$, be defined as the explanation of the set $U_{new}$ in the classifier's latent space.
\begin{equation}
\bar{x} = \frac{1}{k}\sum_{i} x_i
\end{equation}

Being the representation's practical average and let $\delta$ $\epsilon$ (0,1) illustrates,
\begin{equation}
\Delta = \bar{x} - (\frac{v^2}{k} + \frac{k-1}{k}v\delta)^{\frac{1}{2}}>0
\end{equation}

Then the outline is,
\begin{equation}
\text{The equation will be inserted here...}
\end{equation}

Calculated by,
\begin{equation}
\theta \in [\max{\Delta-v,0}, \Delta]
\end{equation}

is a means of solving Problem 2,
\begin{equation}
p_n=\left(1-\frac{C_{new,x}}{2}\left[\rho\left(v^2-(\Delta-\theta)^2\right)^{\frac{1}{2}}\right]^n\right)\times\left(1-C_{new}\left(\frac{k(k-1)}{2}\right)\left[\rho v\left(1-\delta^2\right)^{\frac{1}{2}}\right]^n\right)
\end{equation}

\begin{equation}
p_e = 1 - \frac{C_x}{2}\left[\rho\left(1-\theta^2\right)^{\frac{1}{2}}\right]^n
\end{equation}

Where Cx  and  Cnew are constant.

\begin{longtable}[c]{@{}lllll@{}}
\caption{An illustration of the different methodological approaches, their publishing year, the purpose of the selected study, training and test data type of few-shot learning over 2018 to 2021 in the medical domain.}
\label{table1}\\
\toprule
\textbf{Year} & \textbf{\begin{tabular}[c]{@{}l@{}}Purpose Of \\ the Research\end{tabular}} & \textbf{\begin{tabular}[c]{@{}l@{}}Research \\ Approach\end{tabular}} & \textbf{\begin{tabular}[c]{@{}l@{}}Field Of Training \\ and Testing \\ Domain\end{tabular}} & \textbf{Reference} \\* \midrule
\endfirsthead
\multicolumn{5}{c}%
{{\bfseries Table \thetable\ continued from previous page}} \\
\toprule
\textbf{Year} & \textbf{\begin{tabular}[c]{@{}l@{}}Purpose Of \\ the Research\end{tabular}} & \textbf{\begin{tabular}[c]{@{}l@{}}Research \\ Approach\end{tabular}} & \textbf{\begin{tabular}[c]{@{}l@{}}Field Of Training \\ and Testing \\ Domain\end{tabular}} & \textbf{Reference} \\* \midrule
\endhead
\bottomrule
\endfoot
\endlastfoot
2018 & \begin{tabular}[c]{@{}l@{}}Text Classification\\  on large scale for \\ medical purpose\end{tabular} & \begin{tabular}[c]{@{}l@{}}In this   paper, they used \\ state-of-the-art methods to \\ bring\\ to pass the fine-grained   \\ evaluation. The few and \\ zero-shot \\ methods are evaluated \\ for two \\ medical datasets   \\ Microwave Monolithic \\ Integrated Circuit 2 \\ and Microwave \\ Monolithic Integrated   \\ Circuit 3.\end{tabular} & \begin{tabular}[c]{@{}l@{}}Medical, \\ departure \\ report,\\    \\ medical coding\end{tabular} & \cite{rios2018few} \\
2018 & \begin{tabular}[c]{@{}l@{}}Multi-label\\    \\ Medical coding\\    \\ Classification\end{tabular} & \begin{tabular}[c]{@{}l@{}}A neural network \\ architecture is built \\ from a few-shot \\ learning to match \\ network and   \\ multi-label loss \\ functions. CNN also \\ used to outdo the \\ models. The motive of   \\ this paper is to guide t\\ he EMR dataset for \\ countable multi-label \\ outcomes.\end{tabular} & medical & \cite{rios2018emr} \\
2018 & \begin{tabular}[c]{@{}l@{}}Named entity \\ recognition (NER)\\    \\ for medical records\end{tabular} & \begin{tabular}[c]{@{}l@{}}Assess five developments in \\ NNR to\\  enhance the NLP models. \\ Layer-wise initialization, \\ ]hyperparameter tuning, \\ combining pre-training data, \\ custom word embeddings, \\ and optimizing out of \\ vocabulary.\end{tabular} & medical & \cite{hofer2018few} \\* \midrule
2019 & \begin{tabular}[c]{@{}l@{}}Medical text concept \\ extraction through \\ normalization\end{tabular} & \begin{tabular}[c]{@{}l@{}}Sanctuary for few-shot \\ learning \\ that is built on UMLS data \\ and pre-trained \\ word embeddings. The \\ outsail of this process \\ is the average participant \\ and exhibiting \\ the potential feasibility.\end{tabular} & \begin{tabular}[c]{@{}l@{}}medical (Adverse \\ drug reaction)\end{tabular} & \cite{manousogiannis2019give} \\
2019 & \begin{tabular}[c]{@{}l@{}}Named Entity \\ Recognition (NER)\\ for medical documents\end{tabular} & \begin{tabular}[c]{@{}l@{}}Key phrases in  medical \\ documents are identified \\ and categorized using NER. \\ Hybrid Bi-LSTM   and CNN \\ with four input layers.\end{tabular} & clinical & \cite{lara2019key} \\
2019 & \begin{tabular}[c]{@{}l@{}}Information \\ retrieval task\\ in the biomedical and \\ life science sector \\ through \\ entity normalization\end{tabular} & \begin{tabular}[c]{@{}l@{}}The concept normalization \\ method is used with \\ the amalgamation \\ of distributional semantics, \\ ontological \\ knowledge, and weak \\ supervision \\ which is applied to BB-norm \\ for microbial habitats and \\ phenotypes represented \\ in the text.\end{tabular} & biological & \cite{ferre2020c} \\
2020 & \begin{tabular}[c]{@{}l@{}}Multi-label document \\ classification \\ through \\ graph aggregation \\ model\end{tabular} & \begin{tabular}[c]{@{}l@{}}The paper represents a \\ multi-graph \\ knowledge aggregation \\ model \\ which \\ is trained to combine the s\\ tructural \\ information from multiple \\ labels \\ graphs. The model is used as a \\ sub-module of the previous \\ natural \\ language for few-shot learning.\end{tabular} & medical & \cite{lu2020multi} \\
2020 & \begin{tabular}[c]{@{}l@{}}Collecting data \\ from \\ electronic health \\ records.\end{tabular} & \begin{tabular}[c]{@{}l@{}}A combined model is \\ applied \\ in this paper to extract the \\ salient \\ information from the \\ CRIS dataset \\ for training and testing \\ the statistical \\ model and corporate a \\ fine-tuning \\ model for better efficacy. This \\ method is \\ used in the real project of \\ drug-related \\ information by \\ amalgamating \\ state-of-the-art   \\ active learning paradigms \\ that \\ outperform the existing \\ methods.\end{tabular} & Clinical & \cite{chen2020multimodal} \\
2020 & \begin{tabular}[c]{@{}l@{}}Medical data \\ classification\end{tabular} & \begin{tabular}[c]{@{}l@{}}Based on multi-modal \\ feature fusion \\ and small sample learning \\ techniques, \\ a classification method is \\ applied in \\ this paper for identifying the \\ information \\ of Alzheimer’s patients. \\ The features of \\ this proposed model is\\ extracted \\ at the vector level. Therefore, \\ a small \\ a sample learning \\ network is \\ fabricated by the KNN \\ attention pooling layer and \\ the convolutional \\ network (CNN).\end{tabular} & MRI data & \cite{chen2020multimodal} \\
2020 & \begin{tabular}[c]{@{}l@{}}Sequence tagging\\ through Named Entity \\ Recognition (NER)\end{tabular} & \begin{tabular}[c]{@{}l@{}}The paper is illustrated a \\ prototypical \\ network and relation module \\ for trigger identification to \\ measure \\ the distance of biomedical \\ events. \\ Here, a self-attention \\ mechanism is \\ proposed for using fully \\ complex \\ biomedical knowledge \\ which is known as \\ external knowledge.\end{tabular} & biomedical & \cite{yin2020knowledge} \\
2021 & \begin{tabular}[c]{@{}l@{}}Negation scope\\ resolution in \\ clinical text through \\ concept extraction\end{tabular} & \begin{tabular}[c]{@{}l@{}}The paper represented a \\ method of \\ zero-shot cross-lingual \\ transfer by \\ using negation scope \\ resolution \\ for multilingual information   \\ extraction from clinical texts \\ which is considered a \\ universal \\ approach. This work shows \\ the \\ possibility of a better \\ output not \\ only with the English \\ language \\ but also with Spanish and \\ French \\ clinical texts without \\ dominions of  \\ target training data or language. \\ Therefore, \\ data concentration by exploiting   \\ it and MLT setup are used for \\ better performance.\end{tabular} & clinical & \cite{hartmann2021multilingual} \\
2021 & \begin{tabular}[c]{@{}l@{}}Name normalization of \\ biomedical names by \\ specified domain \\ semantics\end{tabular} & \begin{tabular}[c]{@{}l@{}}A Deep Averaging \\ Network (DAN) \\ model is used \\ to solve the problem of \\ biomedical \\ name representation \\ by aligning encoded \\ names with more\\  domain-specific \\ semantics and conceptual\\ grounding. \\ DAN's \\ feed-forward encoding \\ architecture makes it \\ scalable for use in literal s\\ ynonymy and \\ semantic relatedness tasks, \\ and it has shown \\ expected results in various\\  classifications \\ of biomedical names \\ without requiring \\ LSTM order. The model \\ performs well \\ in categorizations, \\ ontologies, and \\ benchmarks.\end{tabular} & Biomedical & \cite{fivez2021conceptual} \\
2021 & \begin{tabular}[c]{@{}l@{}}Context extraction of \\ molecular feature of \\ drug response prediction\end{tabular} & \begin{tabular}[c]{@{}l@{}}In this study, a flexible neural \\ network model which \\ may be modified for \\ different cell line situations \\ were applied. This system \\ accommodated quickly \\ in case of \\ switching different tissue\\  types \\ and moving from cell line \\ models to the clinical subject \\ and also including \\ patient-derived tumor \\ cells and \\ patient-derived \\ xenografts.\end{tabular} & Biomedical & \cite{ma2021few} \\
2021 & \begin{tabular}[c]{@{}l@{}}Domain knowledge \\ extraction \\ from biomedical entities\end{tabular} & \begin{tabular}[c]{@{}l@{}}The BioGraphSAGE \\ technique \\ removes biological existence \\ relations from literature by \\ using a \\ Siamese graph neural network \\ and structured databases as \\ domain knowledge. \\ The use of the two qualities\\ semantic and positional \\ helps to recognize \\ connections \\ between distant elements \\ in the same literature.\end{tabular} & Biomedical   entities & \cite{guo2021extracting} \\
2021 & \begin{tabular}[c]{@{}l@{}}Retrieving \\ information \\ from \\ biomedical name \\ representation\end{tabular} & \begin{tabular}[c]{@{}l@{}}This paper shows three \\ different \\ approaches, a balanced \\ comparison \\ examination of several \\ exemplary, \\ a modified baseline method, \\ and a \\ the new experimental \\ setting for \\ evaluating \\ the cross-domain \\ generalization \\ ability. \\ When the feature backbone   \\ is shallow, the result states \\ that minimizing intra-class \\ variation is an emerging \\ component.\end{tabular} & Biomedical & \cite{chen2019closer} \\* \bottomrule
\end{longtable}
\section{Future Scope}

Few shot learning is progressing continuously in terms of performance which are represented in previous works. However, the algorithm has some drawbacks in the case of implementation. Here, a well-defined and efficient baseline is significant. The algorithm can be tested with a better algorithm if it can experiment against suitable criteria. Chen et al.~\cite{dhillon2019baseline} and Dhillon et al.~\cite{london2019artificial} proposed picturizing such baselines for differentiating few-shot learning algorithms. Moreover, there need to be further experiments for the improvement of performance for a few shot learning on medical images. However, extensive testing and analysis are required to expect promising results from a few shot learning-used algorithms for use in a practical clinical setup.
The black box architectures of deep networks are not promoted with the change of performance in medical applications~\cite{hayashi2019right}. Few-shot learning will vastly be used for the classification of images, and segmentation is illustrated in~\cite{ge2022few} for diagnosis and validation of medical images; few-shot learning can be a great supporting tool. Furthermore, future works have a wide range of opportunities through the practice session. For flourishing technology, Magnetic Resonance Imaging (MRI) is required a magnetism concept and radio graph technique for analyzing body tissues and structures.
The peaking aim of few-shot learning is to train a model with fewer examples like text-based data humans, semantics complexities, syntax, and structure altogether make it difficult for the model to generalize and learn information, especially with the low number of examples used and evolved exceptional terminologies applied in medical identification texts. There is always a question about the utility of FSL and the area s underexplored~\cite{ge2022few}.

\section{Conclusion}

The recent development and advancement of a few shot learning are represented in this paper. Few shot learning method is applied with limited data to solve the data dependency of deep learning. Especially the problems can be defined and explained through the few-shot learning classification process. Approaches of Few shot learning in executed for NLP in the medical domain. Some approaches to few-shot learning are proposed in various papers, especially classification. General performances of few-shot learning are suitable for real-world application and further research on improving performance. The deep discussion about the few-shot learning application has been assessed by researchers. Finally, future opportunities and applications with limited datasets need to be addressed. This chapter concludes the comprehensive explanation of a few shot learning for the future.
\bibliographystyle{unsrt}  
\bibliography{main}

\end{document}